\documentclass[aps,prl,reprint,superscriptaddress,showpacs]{revtex4-1}

\usepackage[]{hyperref} 
\usepackage{amsmath}
\usepackage{amsfonts}
\usepackage{textcomp}
\usepackage[]{graphicx}
\usepackage{epstopdf}
\usepackage{mathrsfs}
\usepackage{amsbsy}
\usepackage{cleveref}
\usepackage[rightcaption]{sidecap}

\crefformat{equation}{Eq.~(#2#1#3)}
\crefformat{figure}{Fig.~#2#1#3}

\Crefformat{equation}{Equation~(#2#1#3)}
\Crefformat{figure}{Figure~#2#1#3}

\hypersetup{pdfauthor={Wiktor Walasik, Natalia M. Litchinitser},pdftitle={Nonlinear PT PRL}}

\begin{document}

\title{Phase transition in multimode nonlinear parity-time-symmetric waveguide couplers}

\author{Wiktor Walasik}
\affiliation{Department of Electrical Engineering, University at Buffalo, The State University of New York, Buffalo, New York 14260, USA}
\author{Natalia M. Litchinitser}
\affiliation{Department of Electrical Engineering, University at Buffalo, The State University of New York, Buffalo, New York 14260, USA}
\email[]{wiktorwa@buffalo.edu}

\date{\today}

\begin{abstract}
Parity-time-symmetric ($\mathcal{PT}$-symmetric) optical waveguide couplers have become a key component for integrated optics. They offer new possibilities for fast, ultracompact, configurable, all-optical signal processing. Here, we study nonlinear properties of finite-size multimode $\mathcal{PT}$-symmetric couplers and report a peculiar type of dispersion relation in couplers made of more than one dimer. Moreover, we predict a $\mathcal{PT}$~transition triggered by nonlinearity in these structures, and we demonstrate that with the increase of the number of dimers in the system, the transition threshold decreases and converges to the value corresponding to an infinite array. Finally, we present a variety of periodic intensity patterns that can be formed in these couplers depending on the initial excitation. 
\end{abstract}


\maketitle

Quantum mechanical observables correspond to operators with a real spectrum of eigenvalues~\cite{Shankar94}. Hermitian operators are the most well-known class of such operators. However, a more general class of non-Hermitian operators that possess a real spectrum was described by Bender \textit{et al.}~\cite{Bender98,Bender02} in 1998. These parity-time-symmetric ($\mathcal{PT}$-symmetric) operators commute with a subsequent action of the parity operator~$\mathcal{P}$ and the time inversion operator~$\mathcal{T}$. Consequently, $\mathcal{PT}$-symmetric Hamiltonians have real spectra and describe physical phenomena, provided that the complex potentials~$V$ fulfill the condition $V(\mathbf{r}) = V^*(-\mathbf{r})$. Optical systems, where the light propagation is described by the nonlinear Schr\"{o}dinger equation, offer an efficient platform for implementation of $\mathcal{PT}$ symmetry. The complex dielectric permittivity distribution ${\epsilon}(\textbf{r})$ plays the role of potential and the imaginary part of the permittivity corresponds to gain or loss~\cite{El-Ganainy07,Klaiman08}.

The optical $\mathcal{PT}$ symmetry in the linear regime was experimentally demonstrated in waveguide couplers~\cite{Ruter10}, which have a critical importance for the future ultracompact, fast, and configurable all-optical switching. In the nonlinear regime, optical couplers with gain and loss were studied~\cite{Chen92}, where suppression of time reversal~\cite{Sukhorukov10} and unidirectionality~\cite{Ramezani10} were demonstrated. 
Phenomena related to nonlinear $\mathcal{PT}$ symmetry were also investigated in periodic systems, where solitons~\cite{Musslimani08,Suchkov11,Abdullaev11,Alexeeva12,Miri12,Wang13}, breathers~\cite{Barashenkov12} and their stability~\cite{Nixon12,Zezyulin12} were analyzed.

For fixed geometrical parameters, $\mathcal{PT}$-symmetric systems can be in the full or the broken $\mathcal{PT}$-symmetric regime depending on the ratio between the imaginary $\epsilon_{\textrm{IM}}$ and the real part $\epsilon_{\textrm{RE}}$ of the permittivity modulation depth ($\epsilon_{\textrm{IM}}/\epsilon_{\textrm{RE}}$). For low values of this ratio, the system is in the full $\mathcal{PT}$-symmetric regime and has purely real eigenvalues~\cite{Guo09,Ruter10}. 
By increasing the $\epsilon_{\textrm{IM}}/\epsilon_{\textrm{RE}}$ ratio, the $\mathcal{PT}$ transition threshold is reached, and the system transforms into the broken $\mathcal{PT}$-symmetric regime, where a pair of modes (one with gain and one with loss) has  complex conjugate effective indices. 
In the linear regime, the $\epsilon_{\textrm{IM}}/\epsilon_{\textrm{RE}}$ ratio is changed by varying $\epsilon_{\textrm{IM}}$, responsible for gain and loss in the system. However, changes of the real part of the permittivity also influence the ratio of $\epsilon_{\textrm{IM}}/\epsilon_{\textrm{RE}}$. The amplitude of $\epsilon_{\textrm{RE}}$ in nonlinear systems can be controlled by varying the incident light intensity. Recently, a nonlinearity-triggered transition from the full to the broken $\mathcal{PT}$-symmetric regime ($\mathcal{PT}$ transition) was reported in an infinite periodic array of waveguides described by cosine-like permittivity distribution~\cite{Lumer13}.

\begin{figure}[!b]
	\includegraphics[width = 0.60\columnwidth, clip=true, trim = {0 0 0 0}]{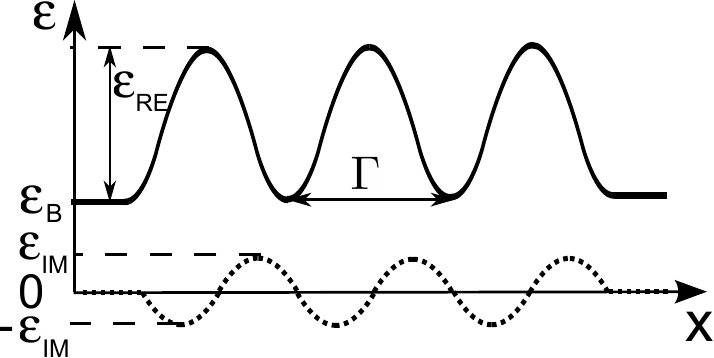}
	\caption{Geometry of the studied $\mathcal{PT}$-symmetric coupler with its parameters. The size (period) of the cosine-like dimers is denoted by $\Gamma$. $\epsilon_B$ denotes the background relative permittivity; $\epsilon_{\textrm{RE}}$ and $\epsilon_{\textrm{IM}}$ denote the modulation amplitude of the real and the imaginary part of relative permittivity, respectively.}
	\label{fig:geom}
\end{figure}

In this paper, we study the nonlinear, $\mathcal{PT}$-symmetric, ultracompact couplers consisting of a few cosine-like waveguide dimers that can be readily integrated on a chip. \Cref{fig:geom} shows the three-dimer geometry under investigation. We report a new type of linear dispersion relation for a multimode $\mathcal{PT}$-symmetric coupler~\cite{Huang14} built of more than one dimer. Additionally, we predict a  $\mathcal{PT}$~transition in such systems and study the dependence of the transition threshold on the number of dimers in the coupler. Peculiar mode interference patterns are reported for the coupler built of three dimers.

Light propagation in one-dimensional (1D) $\mathcal{PT}$-symmetric structures with cubic nonlinearity is studied using a scalar wave equation for the electric field $E(x,z)$:
\begin{equation}
	\left[\nabla^2  + k_0^2 {\epsilon}(x) + k_0^2 \alpha(x) |E|^2\right] E = 0,
	\label{eqn:wave}
\end{equation}
where $k_0 = 2\pi/\lambda$ is the free-space wavevector, $\lambda$ is the free-space wavelength of light, and the operator $\nabla^2 = \partial^2/\partial x^2 + \partial^2/\partial z^2$ denotes the 2D Laplacian. The light propagates along the $z$-direction, and both the structure and the field distributions are assumed to be invariant along the $y$-direction. The linear complex relative permittivity distribution is described by ${\epsilon}(x) = \epsilon_B + \Delta \epsilon(x)$, where $\epsilon_B$ denotes the background relative permittivity and $\Delta \epsilon(x)$ describes the linear complex permittivity modulation depth. The Kerr-type nonlinearity strength is quantified by the parameter $\alpha$. The nonlinearity $\alpha(x)$ is nonzero only inside of the waveguides
.

Using the slowly varying envelope approximation $E(x,z) = \psi(x,z) e^{-i k_0 \sqrt{\epsilon_B} z}$, \cref{eqn:wave} is transformed into the $(1+1)$D nonlinear Schr\"{o}dinger equation
\begin{equation}
\frac{\partial \psi}{\partial z} = -\frac{i}{2 \sqrt{\epsilon_B}} \left[ \frac{1}{k_0} \frac{\partial^2}{\partial x^2} + k_0\left( \Delta {\epsilon} + \alpha |\psi|^2\right)\right] \psi.
\label{eqn:NLSE}
\end{equation}
In order to analyze the nonlinear dynamics of the light propagation in $\mathcal{PT}$-symmetric waveguides, we solve \cref{eqn:NLSE} using the split-step Fourier method \cite{Feit78,Lax81}.

%
%


Previously, nonlinear $\mathcal{PT}$ transition was theoretically studied in an infinite periodic cosine-like waveguide array~\cite{Lumer13}. Here, we analyze simpler, finite-size systems of waveguides, where we find a rich variety of nonlinear phenomena. 
Let us consider a finite-size array built of three dimers (shown in~\cref{fig:geom}) described by 
\begin{equation}
\Delta {\epsilon}(x) =  \epsilon_{\textrm{RE}} \cos^2(\Omega x) + i \epsilon_{\textrm{IM}} \sin(2 \Omega x),
\label{eqn:cos_ind}
\end{equation}
where $\Omega = \pi/\Gamma$, $\Gamma$ denotes the full width of the dimer, and the structure parameters are the following: $\Gamma = 3$~$\mu$m, $\epsilon_{\textrm{RE}} = 0.05$, and $\alpha = 10^{-19}$~m$^2$/V$^2$ at $\lambda = 0.63$~$\mu$m. 

\begin{figure}[!t]
	\includegraphics[width = 0.49\columnwidth, clip=true, trim = {0	0 10 5}]{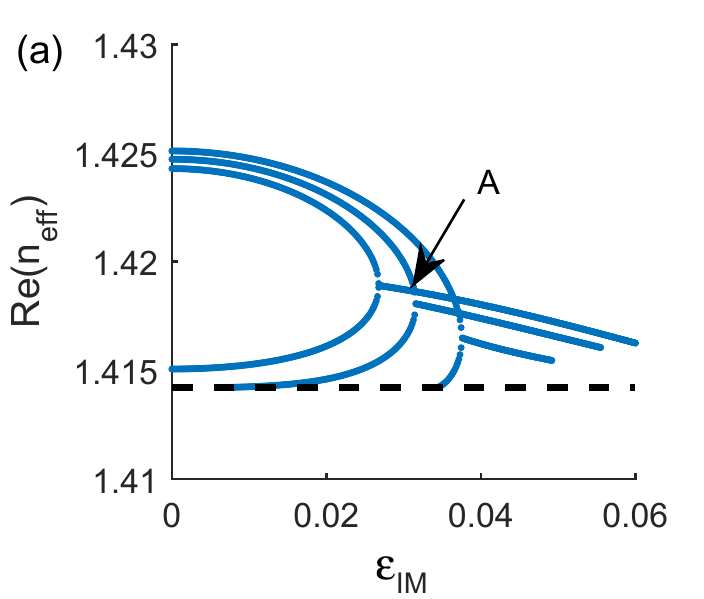}
	\includegraphics[width = 0.49\columnwidth, clip=true, trim = {0 0 10 5}]{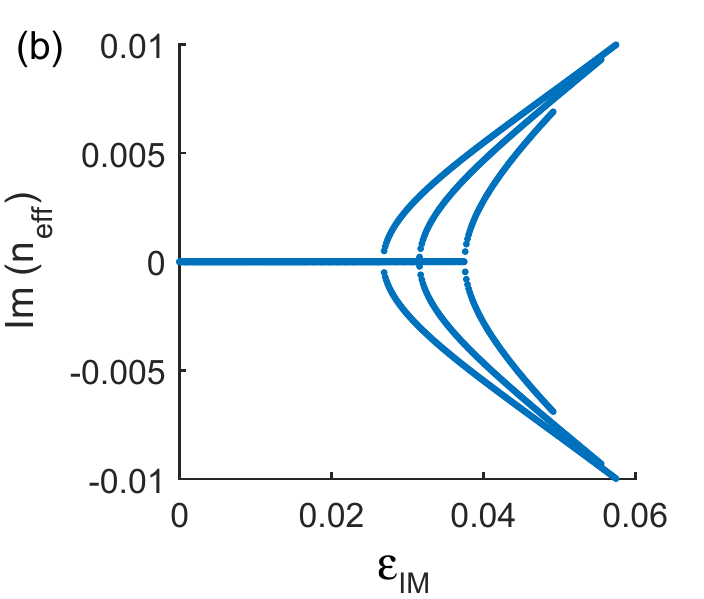}
	\includegraphics[width = 0.49\columnwidth, clip=true, trim = {0 0 10 5}]{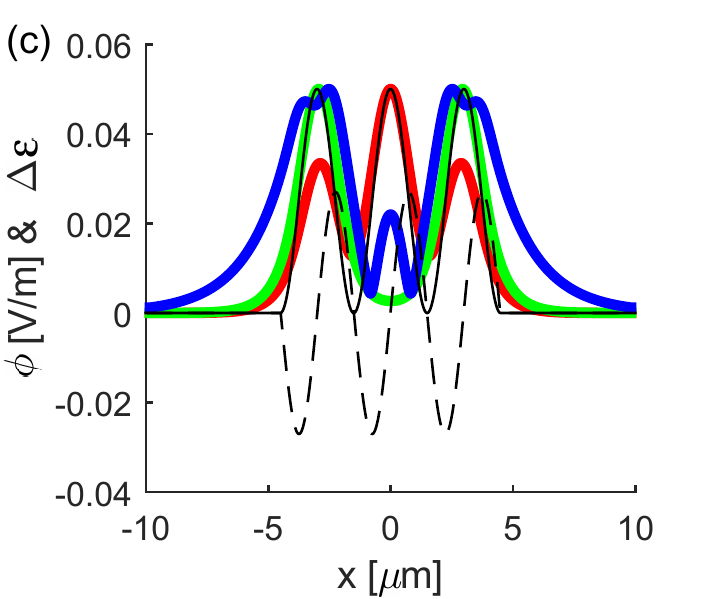}
	\includegraphics[width = 0.49\columnwidth, clip=true, trim = {0 0 10 5}]{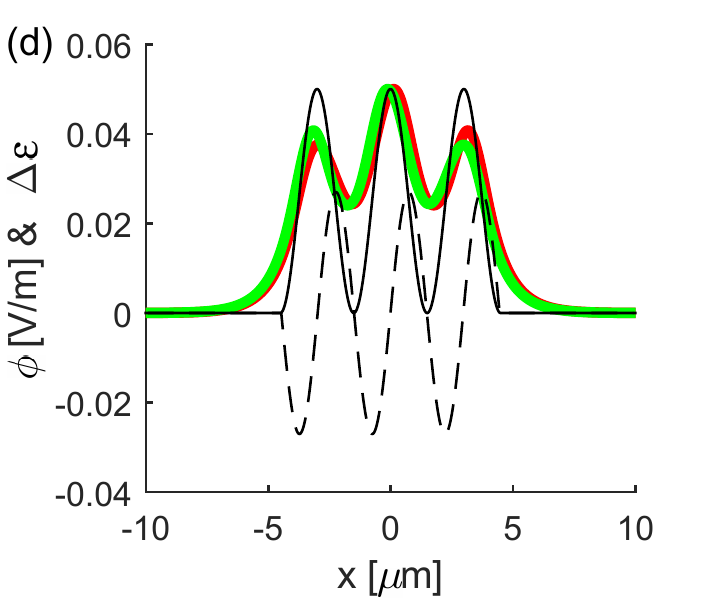}
	\caption{(a),~(b) Dispersion diagrams for a $\mathcal{PT}$-symmetric coupler composed of three dimers: (a) the real and (b) the imaginary part of the effective index of the modes as a function of the imaginary part of permittivity $\epsilon_{\textrm{IM}}$ in the linear case. The black dashed line denotes the refractive index of a background medium $\sqrt{\epsilon_B}$. (c),~(d) Absolute value of the field distributions $|\phi(x)|$ of (c) the lossless modes [mode~1~(fundamental)---red, mode 2---green, mode 5 (lowest effective index)---blue], (d) the coupled modes with loss (green) and gain (red) at $\epsilon_{\textrm{IM}} = 0.027$. Black curves indicate $\Re e\{\Delta \epsilon\}$ (solid) and $\Im m\{\Delta \epsilon\}$ (dashed).}
	\label{fig:disp_multi3}
\end{figure}

Let's first consider a linear system [$\alpha(x) \equiv 0$]. The field profiles that propagate with an effective index $n_{\textrm{eff}}$ without changing their shape are sought in the form $E(x,z) = \phi(x) e^{-i k_0 n_{\textrm{eff}} z}$. This allows us to transform \cref{eqn:wave} into an eigenvalue problem for the modes of the waveguide described by ${\epsilon}(x)$. The eigenvalue problem is solved using a finite-difference method resulting in the dispersion relations [shown in~Figs.~\ref{fig:disp_multi3}(a),~(b)] and the corresponding field profiles of the eigenmodes [shown in~Figs.~\ref{fig:disp_multi3}(c),~(d)]. We use these dispersion diagrams to locate the $\mathcal{PT}$ transition threshold.

For the parameters of the dimers chosen here, the waveguide supports more than one mode. For low values of $\epsilon_{\textrm{IM}}$, there exist four modes with purely real effective indices. With the increase of $\epsilon_{\textrm{IM}}$, two new modes emerge at $\epsilon_{\textrm{IM}} \approx 0.007$ and $0.34$. Meanwhile, the pairs of modes with real effective indices undergo a transition to the broken $\mathcal{PT}$-symmetric regime. For high values of $\epsilon_{\textrm{IM}}$, the higher-order conjugated mode pairs vanish, leaving only the first mode pair from which one mode experiences gain and the other experiences loss during the propagation.

In a single multimode dimer [\cref{fig:geom2}(a) in supplementary material], the dispersion curves of higher-order mode pairs appear below the dispersion curves of the two first modes (see, \cref{fig:single_per_disp_high} in supplementary material and Figs.~2, 5, and 7 in Ref.~\cite{Huang14}).
On the contrary, in the case of an array built of multiple dimers, the dispersion curves of the mode pairs with purely real effective indices do not lay above each other but inside (dispersion curves of the lower-order modes are enclosed by these of the higher-order mode pairs), as it is seen in \cref{fig:disp_multi3}. 
Above the $\mathcal{PT}$~transition, the curve corresponding to $\Re e\{{n_{\textrm{eff}}}\}$ of the two lower-order modes intersects with the dispersion curve of one of the lossless higher-order modes [see e.g., the point labeled $A$ in \cref{fig:disp_multi3}(a) corresponding to $\epsilon_{\textrm{IM}} = 0.0314$]. At these points, three modes with the same value of $\Re e\{{n_{\textrm{eff}}}\}$ exist simultaneously in the dimer array. One of them experiences gain, one loss, and one propagates without change of its amplitude.  

The dispersion relations presented in \cref{fig:disp_multi3} provide the value of $\epsilon_{\textrm{IM}}$ where the first (for lowest value of $\epsilon_{\textrm{IM}}$) $\mathcal{PT}$ threshold occurs in the system. For the array of three dimers, this value is $\epsilon_{\textrm{IM}} = 0.0268$. Below the threshold we find five symmetric (with respect to the center of the array $x=0$) lossles modes with real effective indices. Above the threshold, in addition to the three lossless modes, we find two modes: one for which the field profile is shifted slightly to the gain region (gain mode) and one shifted to the loss region (loss mode). The corresponding field profiles are presented in Figs.~\ref{fig:disp_multi3}(c),~(d).


In addition to the coupler built of three dimers studied here, we have analyzed the $\mathcal{PT}$~transition in structures made of different numbers of dimers. Here, we summarize the values of the $\mathcal{PT}$~transition threshold $\epsilon_{\textrm{IM}}^{(\textrm{th})}$ obtained: for a single dimer $\epsilon_{\textrm{IM}}^{(\textrm{th})} = 0.0305$~\cite{Walasik15}; for two dimers: 0.0278 (see supplementary materials for more details), and for three dimers: 0.0268. Additional calculations show that for the case of four dimers, this value is equal to 0.0262. This suggests two conclusions. Firstly, the threshold value for a finite-size array of dimers is higher than the threshold for the infinite array of dimers described by \cref{eqn:cos_ind}, which was reported in Refs.~\cite{Musslimani08,Lumer13} to be $\epsilon_{\textrm{IM}}^{(\textrm{th})} = 0.5\epsilon_{\textrm{RE}} =  0.025$. This can be understood considering a simple model based on the fact that light concentrates in the regions with high permittivity. Then, the rate with which the overlap integral between a test field profile and the permittivity profile decreases, while shifting the test field profile along the $x$-direction quantifies how difficult it is to break the symmetry in the system and induce the $\mathcal{PT}$~transition. The higher the number of dimers, the slower the change of this overlap integral is and, consequently, the easier it is to induce the $\mathcal{PT}$-symmetry-breaking transition. The second conclusion states that with the increase of the number of dimers, the threshold value converges to the value corresponding to the infinite array as expected because the structure becomes more similar to the infinite array.

\begin{figure}[!t]
	\includegraphics[width = 0.49\columnwidth, clip=true, trim = {0 0 20 0}]{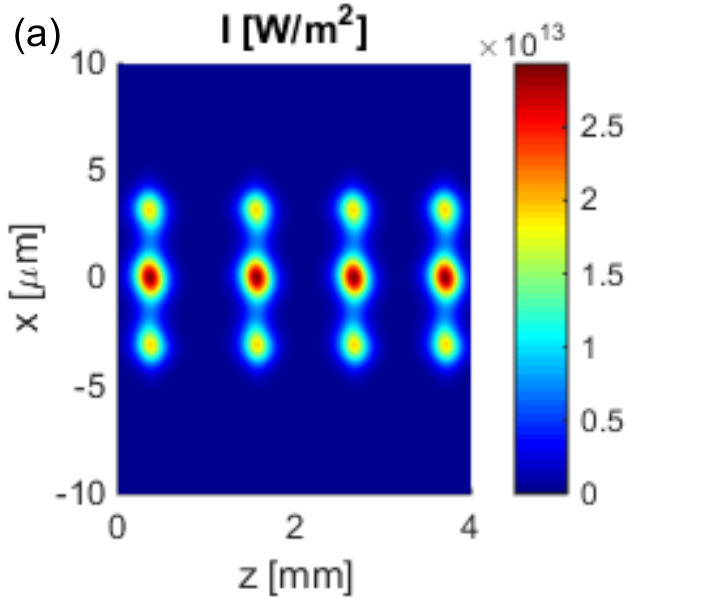}
	\includegraphics[width = 0.49\columnwidth, clip=true, trim = {0 -5 15 5}]{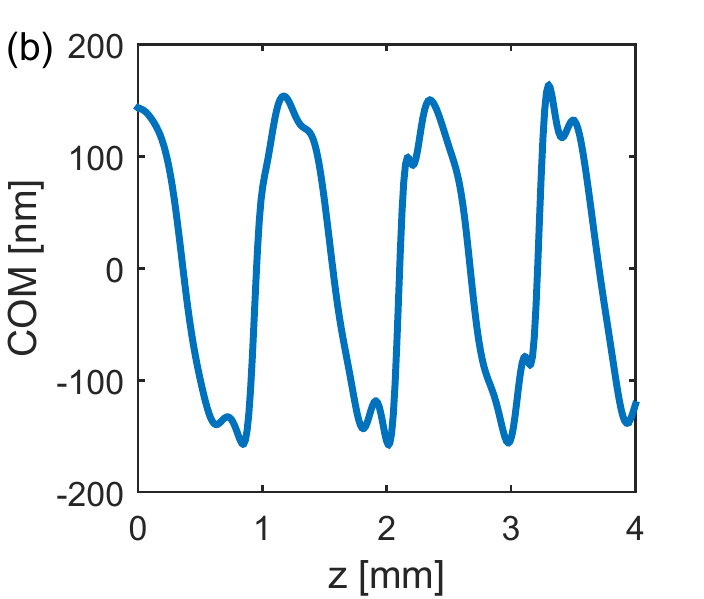}
	\includegraphics[width = 0.49\columnwidth, clip=true, trim = {0 0 20 0}]{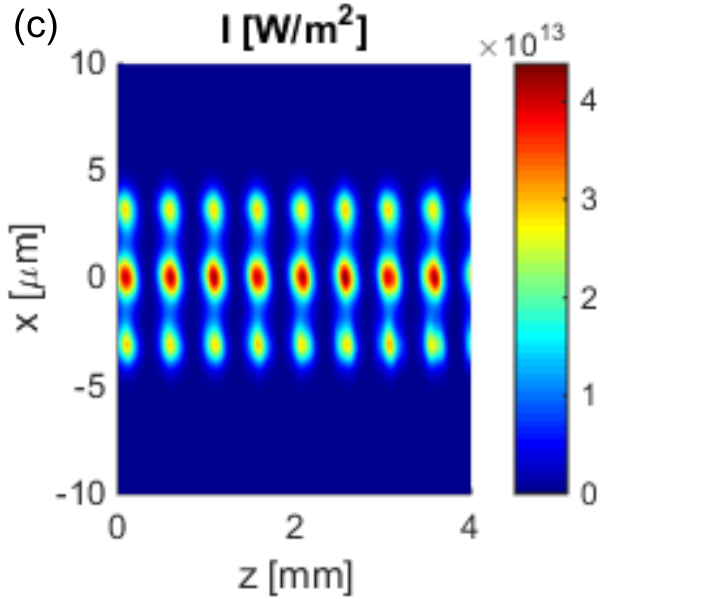}
	\includegraphics[width = 0.49\columnwidth, clip=true, trim = {0 -5 15 5}]{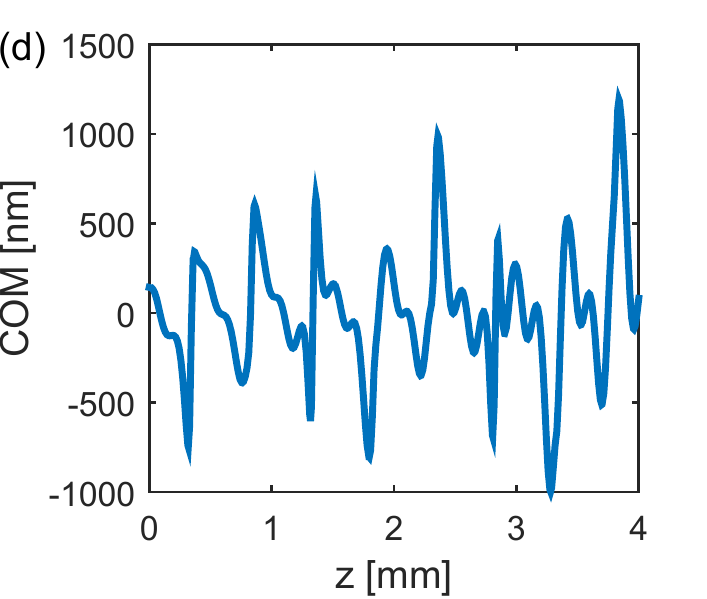}
	\caption{Nonlinear dynamics of light propagating in an array of three dimers described by \cref{eqn:cos_ind} and shown in \cref{fig:geom}. (a),~(c) The intensity distribution $I(x,z)$, (b),~(d) the evolution of the center of mass. $\epsilon_{\textrm{IM}}$ is equal to $0.027$. The input is the linear gain mode with the power density $P_0$ equal to:  (a),~(b) $5\cdot10^6$~W/m and (c),~(d) $10^8$~W/m. }
	\label{fig:three_per_evo}
\end{figure}

\Cref{fig:three_per_evo} presents the evolution of the light propagating in the nonlinear coupler built of three $\mathcal{PT}$-symmetric dimers described by \cref{eqn:cos_ind}. Initially, the system is in the broken $\mathcal{PT}$-symmetric regime, and the light is injected into the gain mode of the system. The intensity of this mode grows rapidly, causing the nonlinear increase of the real part of permittivity, which transforms the system to the full $\mathcal{PT}$-symmetric regime. In this regime, the light is attracted toward the center of each dimer (where the permittivity is the highest), and due to the transverse momentum, light crosses the center and is mostly located in the loss region~\cite{Lumer13}. Consequently, the light intensity and the nonlinear modification of permittivity decrease bringing the system back to the broken $\mathcal{PT}$-symmetric regime. This cycle repeats as shown in~\cref{fig:three_per_evo}(a) and the evolution of the $x$-coordinate of the center of mass (COM) of the intensity distribution $COM(z) = \int_{-\infty}^{+\infty} x I(x,z) \mathrm{d}x/\int_{-\infty}^{+\infty} I(x,z) \mathrm{d}x$ presented in~\cref{fig:three_per_evo}(b). 
For the input beam energy chosen in Figs.~\ref{fig:three_per_evo}(a)~and~(b), the total power in the system increases 25 times during one cycle, and the period of the cycle is approximately equal to 1~mm. Lowering the input power results in an increase of the propagation length required for the power to grow and initiate the oscillations, but the oscillation period and the maximum power remains approximately the same. On the contrary, an increase of the input power results in a decrease of the oscillation period (down to $0.5$~mm) and an increase ($1.5$ times) of the peak power [see Figs.~\ref{fig:three_per_evo}(c),~(d)]. 

\begin{figure}[!t]
	\includegraphics[width = \columnwidth, clip=true, trim = {10 15 35 0}]{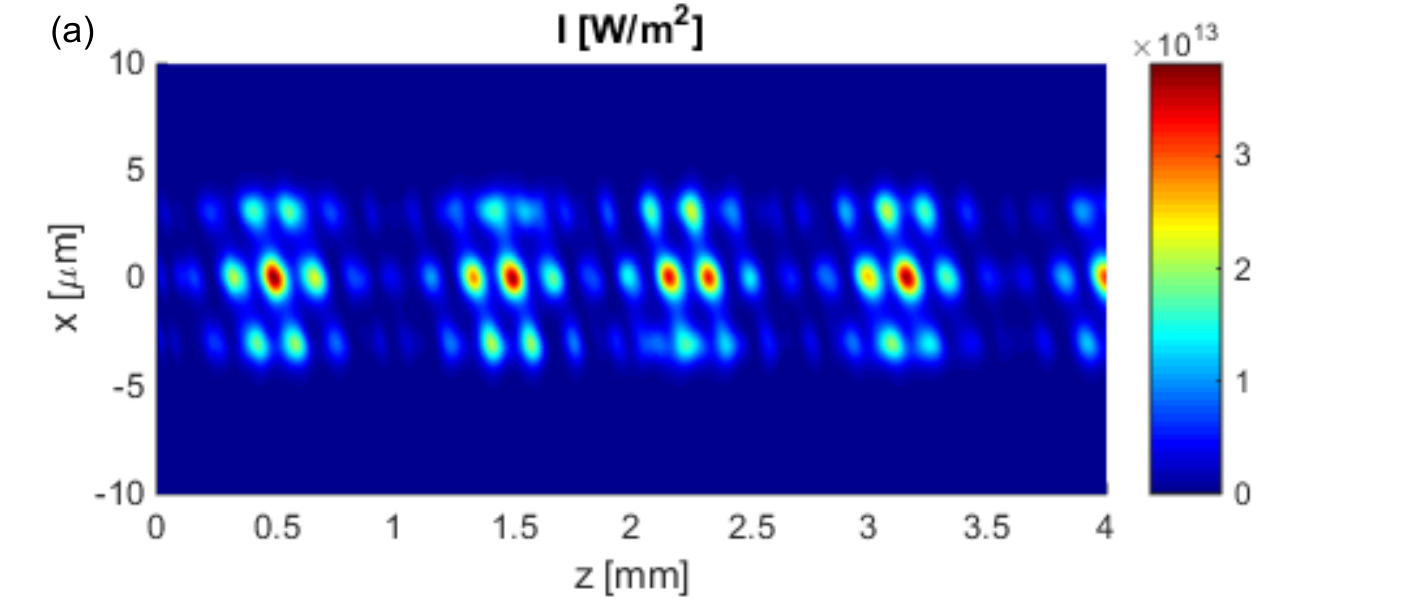}\\
	\includegraphics[width = \columnwidth, clip=true, trim = {10 15 35 5}]{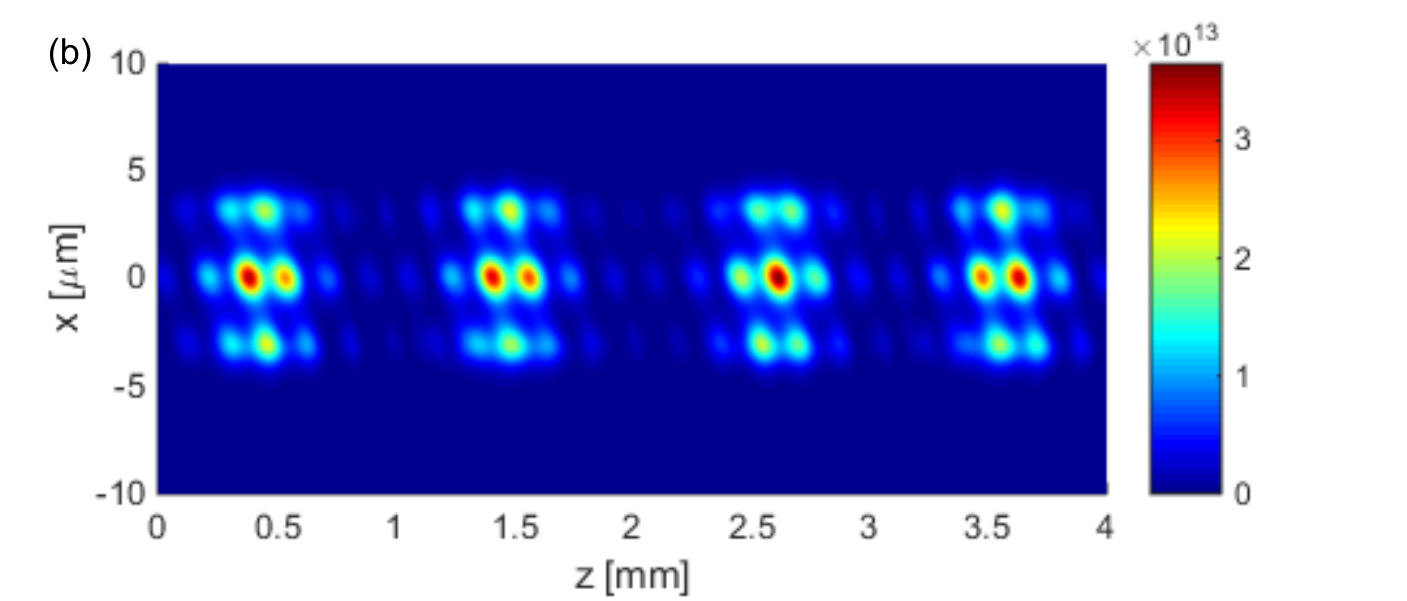}\\
	\includegraphics[width = \columnwidth, clip=true, trim = {10 0 35 5}]{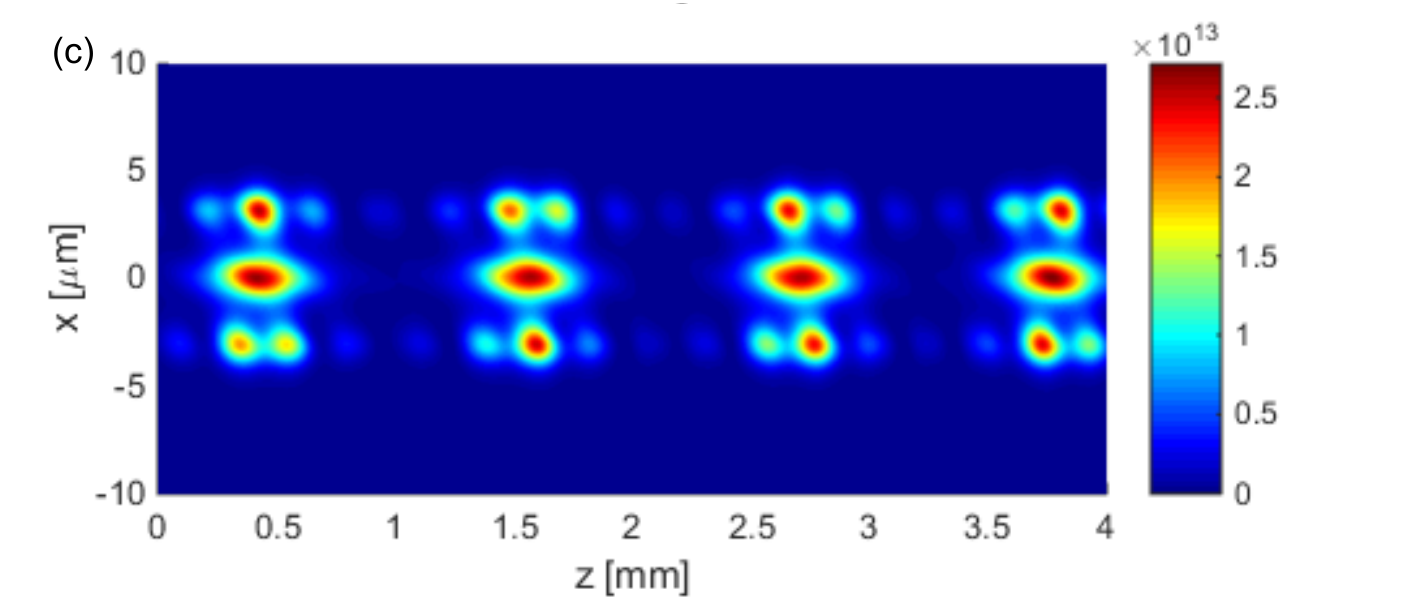}
	\caption{The intensity distributions $I(x,z)$ showing the nonlinear dynamics of light propagating in an array built of three dimers with the same parameters as these presented in \cref{fig:three_per_evo}. The input field is in the form of (a) a wide Gaussian beam illuminating three dimers, (b) the sum of mode~1 and gain mode, (c) the sum of mode~2 and gain mode (for mode numbers and profiles see \cref{fig:disp_multi3}). In all the cases, the excitation power density is $P_0 = 5\cdot10^6$~W/m. }
	\label{fig:three_per_evo_mode12}
\end{figure}

\Cref{fig:three_per_evo_mode12} shows nonlinear propagation of light in the multimode system studied in Figs.~\ref{fig:disp_multi3}~and~\ref{fig:three_per_evo}, but this time instead of exciting a single (gain) mode, we excite a mixture of modes. \Cref{fig:three_per_evo_mode12}(a) shows the result of excitation with a wide Gaussian beam illuminating all three dimers. This beam excites all the modes of the waveguide (except the antisymmetric mode 2), and these modes interfere during the propagation. This interference results in a doubly periodic light distribution pattern. The high frequency oscillations result from the interference of all the modes that have different effective indices. Additionally, we observe low frequency oscillations, that are the result of the nonlinear $\mathcal{PT}$~transition that couples light between the gain and the loss modes. The resulting pattern resembles the triangular lattice. A similar interference pattern is observed when a mixture of the fundamental mode and the gain mode are excited [see \cref{fig:three_per_evo_mode12}(b)].

On the contrary, the excitation of the mixture of mode~2 and the gain mode results in a different interference pattern, as shown in \cref{fig:three_per_evo_mode12}(c). Mode~2 is mainly localized in the side waveguides [see green curves in \cref{fig:disp_multi3}(a)]; therefore, the high frequency interference pattern is observed in these two waveguides only. The low frequency pattern attributed to the energy exchange between the gain and the loss mode is spread over all three waveguide dimers similar to earlier cases. Here, the pattern is less regular than before and consists of an elongated maximum in the middle waveguide and a few shorter maxima in the side waveguides.

In conclusion, we have studied nonlinear dynamics of light propagation in finite-size parity-time-symmetric couplers built of three dimers, and reported a new type of linear dispersion diagram $n_{\textrm{eff}}(\epsilon_{\textrm{IM}})$ for multimode systems built of several dimers, where the dispersion curves of higher-order mode pairs are enclosed in the dispersion curves of lower-order mode pairs. This contrasts with the case of multimode single dimers, where the dispersion curves of higher-order modes appear below the curves corresponding to lower-order modes. Moreover, the nonlinearly triggered transition from the full to the broken $\mathcal{PT}$-symmetric regime can be observed in such a system. The threshold for this $\mathcal{PT}$ transition was found to be higher in finite-size systems than for infinite periodic potential.  Finally, we predicted a variety of periodic patterns can be formed in the multimode coupler by controlling the initial excitation conditions.

\section{Acknowledgements}

Authors thank Dr.~Liang Feng from University at Buffalo, The State University of New York for helpful discussions. This work was supported by US Army Research Office Award \#W911NF-15-1-0152.


\begin{thebibliography}{24}%
	\makeatletter
	\providecommand \@ifxundefined [1]{%
		\@ifx{#1\undefined}
	}%
	\providecommand \@ifnum [1]{%
		\ifnum #1\expandafter \@firstoftwo
		\else \expandafter \@secondoftwo
		\fi
	}%
	\providecommand \@ifx [1]{%
		\ifx #1\expandafter \@firstoftwo
		\else \expandafter \@secondoftwo
		\fi
	}%
	\providecommand \natexlab [1]{#1}%
	\providecommand \enquote  [1]{``#1''}%
	\providecommand \bibnamefont  [1]{#1}%
	\providecommand \bibfnamefont [1]{#1}%
	\providecommand \citenamefont [1]{#1}%
	\providecommand \href@noop [0]{\@secondoftwo}%
	\providecommand \href [0]{\begingroup \@sanitize@url \@href}%
	\providecommand \@href[1]{\@@startlink{#1}\@@href}%
	\providecommand \@@href[1]{\endgroup#1\@@endlink}%
	\providecommand \@sanitize@url [0]{\catcode `\\12\catcode `\$12\catcode
		`\&12\catcode `\#12\catcode `\^12\catcode `\_12\catcode `\%12\relax}%
	\providecommand \@@startlink[1]{}%
	\providecommand \@@endlink[0]{}%
	\providecommand \url  [0]{\begingroup\@sanitize@url \@url }%
	\providecommand \@url [1]{\endgroup\@href {#1}{\urlprefix }}%
	\providecommand \urlprefix  [0]{URL }%
	\providecommand \Eprint [0]{\href }%
	\providecommand \doibase [0]{http://dx.doi.org/}%
	\providecommand \selectlanguage [0]{\@gobble}%
	\providecommand \bibinfo  [0]{\@secondoftwo}%
	\providecommand \bibfield  [0]{\@secondoftwo}%
	\providecommand \translation [1]{[#1]}%
	\providecommand \BibitemOpen [0]{}%
	\providecommand \bibitemStop [0]{}%
	\providecommand \bibitemNoStop [0]{.\EOS\space}%
	\providecommand \EOS [0]{\spacefactor3000\relax}%
	\providecommand \BibitemShut  [1]{\csname bibitem#1\endcsname}%
	\let\auto@bib@innerbib\@empty
	\bibitem [{\citenamefont {Shankar}(1994)}]{Shankar94}%
	\BibitemOpen
	\bibfield  {author} {\bibinfo {author} {\bibfnamefont {R.}~\bibnamefont
			{Shankar}},\ }\href@noop {} {\emph {\bibinfo {title} {Principles of Quantum
				Mechanics, $2^{nd}$ ed.}}}\ (\bibinfo  {publisher} {Plenum Press, New York},\
	\bibinfo {year} {1994})\BibitemShut {NoStop}%
	\bibitem [{\citenamefont {Bender}\ and\ \citenamefont
		{Bottcher}(1998)}]{Bender98}%
	\BibitemOpen
	\bibfield  {author} {\bibinfo {author} {\bibfnamefont {C.~M.}\ \bibnamefont
			{Bender}}\ and\ \bibinfo {author} {\bibfnamefont {S.}~\bibnamefont
			{Bottcher}},\ }\href@noop {} {\bibfield  {journal} {\bibinfo  {journal}
			{Phys. Rev. Lett.}\ }\textbf {\bibinfo {volume} {80}},\ \bibinfo {pages}
		{5243} (\bibinfo {year} {1998})}\BibitemShut {NoStop}%
	\bibitem [{\citenamefont {Bender}\ \emph {et~al.}(2002)\citenamefont {Bender},
		\citenamefont {Brody},\ and\ \citenamefont {Jones}}]{Bender02}%
	\BibitemOpen
	\bibfield  {author} {\bibinfo {author} {\bibfnamefont {C.~M.}\ \bibnamefont
			{Bender}}, \bibinfo {author} {\bibfnamefont {D.~C.}\ \bibnamefont {Brody}}, \
		and\ \bibinfo {author} {\bibfnamefont {H.~F.}\ \bibnamefont {Jones}},\
	}\href@noop {} {\bibfield  {journal} {\bibinfo  {journal} {Phys. Rev. Lett.}\
	}\textbf {\bibinfo {volume} {89}},\ \bibinfo {pages} {270401} (\bibinfo
	{year} {2002})}\BibitemShut {NoStop}%
\bibitem [{\citenamefont {El-Ganainy}\ \emph {et~al.}(2007)\citenamefont
	{El-Ganainy}, \citenamefont {Makris}, \citenamefont {Christodoulides},\ and\
	\citenamefont {Musslimani}}]{El-Ganainy07}%
\BibitemOpen
\bibfield  {author} {\bibinfo {author} {\bibfnamefont {R.}~\bibnamefont
		{El-Ganainy}}, \bibinfo {author} {\bibfnamefont {K.~G.}\ \bibnamefont
		{Makris}}, \bibinfo {author} {\bibfnamefont {D.~N.}\ \bibnamefont
		{Christodoulides}}, \ and\ \bibinfo {author} {\bibfnamefont {Z.~H.}\
		\bibnamefont {Musslimani}},\ }\href@noop {} {\bibfield  {journal} {\bibinfo
		{journal} {Opt. Lett.}\ }\textbf {\bibinfo {volume} {32}},\ \bibinfo {pages}
	{2632} (\bibinfo {year} {2007})}\BibitemShut {NoStop}%
\bibitem [{\citenamefont {Klaiman}\ \emph {et~al.}(2008)\citenamefont
	{Klaiman}, \citenamefont {G\"unther},\ and\ \citenamefont
	{Moiseyev}}]{Klaiman08}%
\BibitemOpen
\bibfield  {author} {\bibinfo {author} {\bibfnamefont {S.}~\bibnamefont
		{Klaiman}}, \bibinfo {author} {\bibfnamefont {U.}~\bibnamefont {G\"unther}},
	\ and\ \bibinfo {author} {\bibfnamefont {N.}~\bibnamefont {Moiseyev}},\
}\href@noop {} {\bibfield  {journal} {\bibinfo  {journal} {Phys. Rev. Lett.}\
}\textbf {\bibinfo {volume} {101}},\ \bibinfo {pages} {080402} (\bibinfo
{year} {2008})}\BibitemShut {NoStop}%
\bibitem [{\citenamefont {R\"uter}\ \emph {et~al.}(2010)\citenamefont
	{R\"uter}, \citenamefont {Makris}, \citenamefont {El-Ganainy}, \citenamefont
	{Christodoulides}, \citenamefont {Segev},\ and\ \citenamefont
	{Kip}}]{Ruter10}%
\BibitemOpen
\bibfield  {author} {\bibinfo {author} {\bibfnamefont {C.~E.}\ \bibnamefont
		{R\"uter}}, \bibinfo {author} {\bibfnamefont {K.~G.}\ \bibnamefont {Makris}},
	\bibinfo {author} {\bibfnamefont {R.}~\bibnamefont {El-Ganainy}}, \bibinfo
	{author} {\bibfnamefont {D.~N.}\ \bibnamefont {Christodoulides}}, \bibinfo
	{author} {\bibfnamefont {M.}~\bibnamefont {Segev}}, \ and\ \bibinfo {author}
	{\bibfnamefont {D.}~\bibnamefont {Kip}},\ }\href@noop {} {\bibfield
	{journal} {\bibinfo  {journal} {Nature Phys.}\ }\textbf {\bibinfo {volume}
		{6}},\ \bibinfo {pages} {192} (\bibinfo {year} {2010})}\BibitemShut {NoStop}%
\bibitem [{\citenamefont {Chen}\ \emph {et~al.}(1992)\citenamefont {Chen},
	\citenamefont {Snyder},\ and\ \citenamefont {Payne}}]{Chen92}%
\BibitemOpen
\bibfield  {author} {\bibinfo {author} {\bibfnamefont {Y.}~\bibnamefont
		{Chen}}, \bibinfo {author} {\bibfnamefont {A.~W.}\ \bibnamefont {Snyder}}, \
	and\ \bibinfo {author} {\bibfnamefont {D.~N.}\ \bibnamefont {Payne}},\
}\href@noop {} {\bibfield  {journal} {\bibinfo  {journal} {{IEEE} Journ.
		Quant. Electron.}\ }\textbf {\bibinfo {volume} {28}},\ \bibinfo {pages} {239}
(\bibinfo {year} {1992})}\BibitemShut {NoStop}%
\bibitem [{\citenamefont {Sukhorukov}\ \emph {et~al.}(2010)\citenamefont
	{Sukhorukov}, \citenamefont {Xu},\ and\ \citenamefont
	{Kivshar}}]{Sukhorukov10}%
\BibitemOpen
\bibfield  {author} {\bibinfo {author} {\bibfnamefont {A.~A.}\ \bibnamefont
		{Sukhorukov}}, \bibinfo {author} {\bibfnamefont {Z.}~\bibnamefont {Xu}}, \
	and\ \bibinfo {author} {\bibfnamefont {Y.~S.}\ \bibnamefont {Kivshar}},\
}\href@noop {} {\bibfield  {journal} {\bibinfo  {journal} {Phys. Rev. A}\
}\textbf {\bibinfo {volume} {82}},\ \bibinfo {pages} {043818} (\bibinfo
{year} {2010})}\BibitemShut {NoStop}%
\bibitem [{\citenamefont {Ramezani}\ \emph {et~al.}(2010)\citenamefont
	{Ramezani}, \citenamefont {Kottos}, \citenamefont {El-Ganainy},\ and\
	\citenamefont {Christodoulides}}]{Ramezani10}%
\BibitemOpen
\bibfield  {author} {\bibinfo {author} {\bibfnamefont {H.}~\bibnamefont
		{Ramezani}}, \bibinfo {author} {\bibfnamefont {T.}~\bibnamefont {Kottos}},
	\bibinfo {author} {\bibfnamefont {R.}~\bibnamefont {El-Ganainy}}, \ and\
	\bibinfo {author} {\bibfnamefont {D.~N.}\ \bibnamefont {Christodoulides}},\
}\href@noop {} {\bibfield  {journal} {\bibinfo  {journal} {Phys. Rev. A}\
}\textbf {\bibinfo {volume} {82}},\ \bibinfo {pages} {043803} (\bibinfo
{year} {2010})}\BibitemShut {NoStop}%
\bibitem [{\citenamefont {Musslimani}\ \emph {et~al.}(2008)\citenamefont
	{Musslimani}, \citenamefont {Makris}, \citenamefont {El-Ganainy},\ and\
	\citenamefont {Christodoulides}}]{Musslimani08}%
\BibitemOpen
\bibfield  {author} {\bibinfo {author} {\bibfnamefont {Z.~H.}\ \bibnamefont
		{Musslimani}}, \bibinfo {author} {\bibfnamefont {K.~G.}\ \bibnamefont
		{Makris}}, \bibinfo {author} {\bibfnamefont {R.}~\bibnamefont {El-Ganainy}},
	\ and\ \bibinfo {author} {\bibfnamefont {D.~N.}\ \bibnamefont
		{Christodoulides}},\ }\href@noop {} {\bibfield  {journal} {\bibinfo
		{journal} {Phys. Rev. Lett.}\ }\textbf {\bibinfo {volume} {100}},\ \bibinfo
	{pages} {030402} (\bibinfo {year} {2008})}\BibitemShut {NoStop}%
\bibitem [{\citenamefont {Suchkov}\ \emph {et~al.}(2011)\citenamefont
	{Suchkov}, \citenamefont {Malomed}, \citenamefont {Dmitriev},\ and\
	\citenamefont {Kivshar}}]{Suchkov11}%
\BibitemOpen
\bibfield  {author} {\bibinfo {author} {\bibfnamefont {S.~V.}\ \bibnamefont
		{Suchkov}}, \bibinfo {author} {\bibfnamefont {B.~A.}\ \bibnamefont
		{Malomed}}, \bibinfo {author} {\bibfnamefont {S.~V.}\ \bibnamefont
		{Dmitriev}}, \ and\ \bibinfo {author} {\bibfnamefont {Y.~S.}\ \bibnamefont
		{Kivshar}},\ }\href@noop {} {\bibfield  {journal} {\bibinfo  {journal} {Phys.
			Rev. E}\ }\textbf {\bibinfo {volume} {84}},\ \bibinfo {pages} {046609}
	(\bibinfo {year} {2011})}\BibitemShut {NoStop}%
\bibitem [{\citenamefont {Abdullaev}\ \emph {et~al.}(2011)\citenamefont
	{Abdullaev}, \citenamefont {Kartashov}, \citenamefont {Konotop},\ and\
	\citenamefont {Zezyulin}}]{Abdullaev11}%
\BibitemOpen
\bibfield  {author} {\bibinfo {author} {\bibfnamefont {F.~K.}\ \bibnamefont
		{Abdullaev}}, \bibinfo {author} {\bibfnamefont {Y.~V.}\ \bibnamefont
		{Kartashov}}, \bibinfo {author} {\bibfnamefont {V.~V.}\ \bibnamefont
		{Konotop}}, \ and\ \bibinfo {author} {\bibfnamefont {D.~A.}\ \bibnamefont
		{Zezyulin}},\ }\href@noop {} {\bibfield  {journal} {\bibinfo  {journal}
		{Phys. Rev. A}\ }\textbf {\bibinfo {volume} {83}},\ \bibinfo {pages} {041805}
	(\bibinfo {year} {2011})}\BibitemShut {NoStop}%
\bibitem [{\citenamefont {Alexeeva}\ \emph {et~al.}(2012)\citenamefont
	{Alexeeva}, \citenamefont {Barashenkov}, \citenamefont {Sukhorukov},\ and\
	\citenamefont {Kivshar}}]{Alexeeva12}%
\BibitemOpen
\bibfield  {author} {\bibinfo {author} {\bibfnamefont {N.~V.}\ \bibnamefont
		{Alexeeva}}, \bibinfo {author} {\bibfnamefont {I.~V.}\ \bibnamefont
		{Barashenkov}}, \bibinfo {author} {\bibfnamefont {A.~A.}\ \bibnamefont
		{Sukhorukov}}, \ and\ \bibinfo {author} {\bibfnamefont {Y.~S.}\ \bibnamefont
		{Kivshar}},\ }\href@noop {} {\bibfield  {journal} {\bibinfo  {journal} {Phys.
			Rev. A}\ }\textbf {\bibinfo {volume} {85}},\ \bibinfo {pages} {063837}
	(\bibinfo {year} {2012})}\BibitemShut {NoStop}%
\bibitem [{\citenamefont {Miri}\ \emph {et~al.}(2012)\citenamefont {Miri},
	\citenamefont {Aceves}, \citenamefont {Kottos}, \citenamefont {Kovanis},\
	and\ \citenamefont {Christodoulides}}]{Miri12}%
\BibitemOpen
\bibfield  {author} {\bibinfo {author} {\bibfnamefont {M.-A.}\ \bibnamefont
		{Miri}}, \bibinfo {author} {\bibfnamefont {A.~B.}\ \bibnamefont {Aceves}},
	\bibinfo {author} {\bibfnamefont {T.}~\bibnamefont {Kottos}}, \bibinfo
	{author} {\bibfnamefont {V.}~\bibnamefont {Kovanis}}, \ and\ \bibinfo
	{author} {\bibfnamefont {D.~N.}\ \bibnamefont {Christodoulides}},\
}\href@noop {} {\bibfield  {journal} {\bibinfo  {journal} {Phys. Rev. A}\
}\textbf {\bibinfo {volume} {86}},\ \bibinfo {pages} {033801} (\bibinfo
{year} {2012})}\BibitemShut {NoStop}%
\bibitem [{\citenamefont {Wang}\ and\ \citenamefont {Aceves}(2013)}]{Wang13}%
\BibitemOpen
\bibfield  {author} {\bibinfo {author} {\bibfnamefont {D.}~\bibnamefont
		{Wang}}\ and\ \bibinfo {author} {\bibfnamefont {A.~B.}\ \bibnamefont
		{Aceves}},\ }\href@noop {} {\bibfield  {journal} {\bibinfo  {journal} {Phys.
			Rev. A}\ }\textbf {\bibinfo {volume} {88}},\ \bibinfo {pages} {043831}
	(\bibinfo {year} {2013})}\BibitemShut {NoStop}%
\bibitem [{\citenamefont {Barashenkov}\ \emph {et~al.}(2012)\citenamefont
	{Barashenkov}, \citenamefont {Suchkov}, \citenamefont {Sukhorukov},
	\citenamefont {Dmitriev},\ and\ \citenamefont {Kivshar}}]{Barashenkov12}%
\BibitemOpen
\bibfield  {author} {\bibinfo {author} {\bibfnamefont {I.~V.}\ \bibnamefont
		{Barashenkov}}, \bibinfo {author} {\bibfnamefont {S.~V.}\ \bibnamefont
		{Suchkov}}, \bibinfo {author} {\bibfnamefont {A.~A.}\ \bibnamefont
		{Sukhorukov}}, \bibinfo {author} {\bibfnamefont {S.~V.}\ \bibnamefont
		{Dmitriev}}, \ and\ \bibinfo {author} {\bibfnamefont {Y.~S.}\ \bibnamefont
		{Kivshar}},\ }\href@noop {} {\bibfield  {journal} {\bibinfo  {journal} {Phys.
			Rev. A}\ }\textbf {\bibinfo {volume} {86}},\ \bibinfo {pages} {053809}
	(\bibinfo {year} {2012})}\BibitemShut {NoStop}%
\bibitem [{\citenamefont {Nixon}\ \emph {et~al.}(2012)\citenamefont {Nixon},
	\citenamefont {Ge},\ and\ \citenamefont {Yang}}]{Nixon12}%
\BibitemOpen
\bibfield  {author} {\bibinfo {author} {\bibfnamefont {S.}~\bibnamefont
		{Nixon}}, \bibinfo {author} {\bibfnamefont {L.}~\bibnamefont {Ge}}, \ and\
	\bibinfo {author} {\bibfnamefont {J.}~\bibnamefont {Yang}},\ }\href@noop {}
{\bibfield  {journal} {\bibinfo  {journal} {Phys. Rev. A}\ }\textbf {\bibinfo
		{volume} {85}},\ \bibinfo {pages} {023822} (\bibinfo {year}
	{2012})}\BibitemShut {NoStop}%
\bibitem [{\citenamefont {Zezyulin}\ and\ \citenamefont
	{Konotop}(2012)}]{Zezyulin12}%
\BibitemOpen
\bibfield  {author} {\bibinfo {author} {\bibfnamefont {D.~A.}\ \bibnamefont
		{Zezyulin}}\ and\ \bibinfo {author} {\bibfnamefont {V.~V.}\ \bibnamefont
		{Konotop}},\ }\href@noop {} {\bibfield  {journal} {\bibinfo  {journal} {Phys.
			Rev. Lett.}\ }\textbf {\bibinfo {volume} {108}},\ \bibinfo {pages} {213906}
	(\bibinfo {year} {2012})}\BibitemShut {NoStop}%
\bibitem [{\citenamefont {Guo}\ \emph {et~al.}(2009)\citenamefont {Guo},
	\citenamefont {Salamo}, \citenamefont {Duchesne}, \citenamefont {Morandotti},
	\citenamefont {Volatier-Ravat}, \citenamefont {Aimez}, \citenamefont
	{Siviloglou},\ and\ \citenamefont {Christodoulides}}]{Guo09}%
\BibitemOpen
\bibfield  {author} {\bibinfo {author} {\bibfnamefont {A.}~\bibnamefont
		{Guo}}, \bibinfo {author} {\bibfnamefont {G.~J.}\ \bibnamefont {Salamo}},
	\bibinfo {author} {\bibfnamefont {D.}~\bibnamefont {Duchesne}}, \bibinfo
	{author} {\bibfnamefont {R.}~\bibnamefont {Morandotti}}, \bibinfo {author}
	{\bibfnamefont {M.}~\bibnamefont {Volatier-Ravat}}, \bibinfo {author}
	{\bibfnamefont {V.}~\bibnamefont {Aimez}}, \bibinfo {author} {\bibfnamefont
		{G.~A.}\ \bibnamefont {Siviloglou}}, \ and\ \bibinfo {author} {\bibfnamefont
		{D.~N.}\ \bibnamefont {Christodoulides}},\ }\href@noop {} {\bibfield
	{journal} {\bibinfo  {journal} {Phys. Rev. Lett.}\ }\textbf {\bibinfo
		{volume} {103}},\ \bibinfo {pages} {093902} (\bibinfo {year}
	{2009})}\BibitemShut {NoStop}%
\bibitem [{\citenamefont {Lumer}\ \emph {et~al.}(2013)\citenamefont {Lumer},
	\citenamefont {Plotnik}, \citenamefont {Rechtsman},\ and\ \citenamefont
	{Segev}}]{Lumer13}%
\BibitemOpen
\bibfield  {author} {\bibinfo {author} {\bibfnamefont {Y.}~\bibnamefont
		{Lumer}}, \bibinfo {author} {\bibfnamefont {Y.}~\bibnamefont {Plotnik}},
	\bibinfo {author} {\bibfnamefont {M.~C.}\ \bibnamefont {Rechtsman}}, \ and\
	\bibinfo {author} {\bibfnamefont {M.}~\bibnamefont {Segev}},\ }\href@noop {}
{\bibfield  {journal} {\bibinfo  {journal} {Phys. Rev. Lett.}\ }\textbf
	{\bibinfo {volume} {111}},\ \bibinfo {pages} {263901} (\bibinfo {year}
	{2013})}\BibitemShut {NoStop}%
\bibitem [{\citenamefont {Huang}\ \emph {et~al.}(2014)\citenamefont {Huang},
	\citenamefont {Ye},\ and\ \citenamefont {Chen}}]{Huang14}%
\BibitemOpen
\bibfield  {author} {\bibinfo {author} {\bibfnamefont {C.}~\bibnamefont
		{Huang}}, \bibinfo {author} {\bibfnamefont {F.}~\bibnamefont {Ye}}, \ and\
	\bibinfo {author} {\bibfnamefont {X.}~\bibnamefont {Chen}},\ }\href@noop {}
{\bibfield  {journal} {\bibinfo  {journal} {Phys. Rev. A}\ }\textbf {\bibinfo
		{volume} {90}},\ \bibinfo {pages} {043833} (\bibinfo {year}
	{2014})}\BibitemShut {NoStop}%
\bibitem [{\citenamefont {Feit}\ and\ \citenamefont {Fleck{,
			Jr.}}(1978)}]{Feit78}%
\BibitemOpen
\bibfield  {author} {\bibinfo {author} {\bibfnamefont {M.~D.}\ \bibnamefont
		{Feit}}\ and\ \bibinfo {author} {\bibfnamefont {J.~A.}\ \bibnamefont {Fleck{,
				Jr.}}},\ }\href@noop {} {\bibfield  {journal} {\bibinfo  {journal} {Appl.
			Opt.}\ }\textbf {\bibinfo {volume} {17}},\ \bibinfo {pages} {3990} (\bibinfo
	{year} {1978})}\BibitemShut {NoStop}%
\bibitem [{\citenamefont {Lax}\ \emph {et~al.}(1981)\citenamefont {Lax},
	\citenamefont {Batteh},\ and\ \citenamefont {Agrawal}}]{Lax81}%
\BibitemOpen
\bibfield  {author} {\bibinfo {author} {\bibfnamefont {M.}~\bibnamefont
		{Lax}}, \bibinfo {author} {\bibfnamefont {J.~H.}\ \bibnamefont {Batteh}}, \
	and\ \bibinfo {author} {\bibfnamefont {G.~P.}\ \bibnamefont {Agrawal}},\
}\href@noop {} {\bibfield  {journal} {\bibinfo  {journal} {J. Appl. Phys.}\
}\textbf {\bibinfo {volume} {52}},\ \bibinfo {pages} {109} (\bibinfo {year}
{1981})}\BibitemShut {NoStop}%
\bibitem [{\citenamefont {Walasik}\ \emph {et~al.}(2015)\citenamefont
	{Walasik}, \citenamefont {Ma},\ and\ \citenamefont
	{Litchinitser}}]{Walasik15}%
\BibitemOpen
\bibfield  {author} {\bibinfo {author} {\bibfnamefont {W.}~\bibnamefont
		{Walasik}}, \bibinfo {author} {\bibfnamefont {C.}~\bibnamefont {Ma}}, \ and\
	\bibinfo {author} {\bibfnamefont {N.~M.}\ \bibnamefont {Litchinitser}},\
}\href@noop {} {\bibfield  {journal} {\bibinfo  {journal} {ArXiv}\ }
(\bibinfo {year} {2015})}\BibitemShut {NoStop}%
\end{thebibliography}
%

\clearpage

\appendix

\renewcommand{\thefigure}{S\arabic{figure}}

\setcounter{figure}{0}

\section{Supplementary materials}

\begin{figure}[!h]
	\includegraphics[width = \columnwidth, clip=true, trim = {0 0 0 0}]{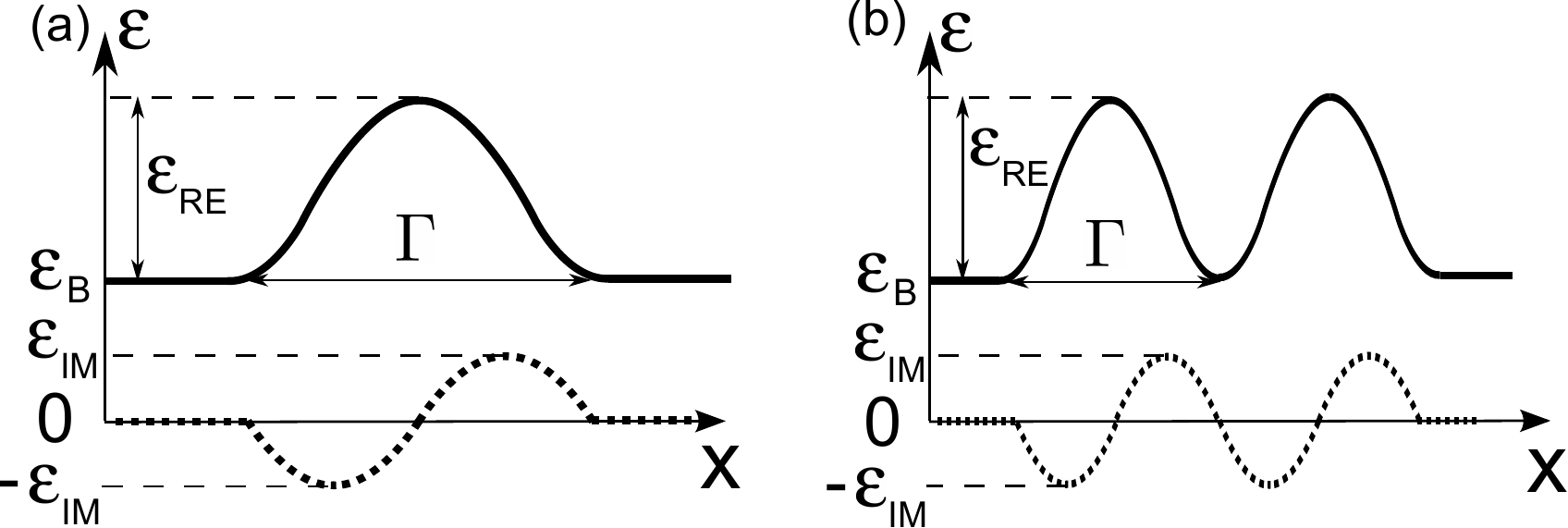}
	\caption{Geometry of the $\mathcal{PT}$-symmetric couplers built of (a) one and (b) two cosine-like dimers. The period of the cosine-like dimers is denoted by $\Gamma$. $\epsilon_B$ denotes the background relative permittivity; $\epsilon_{\textrm{RE}}$ and $\epsilon_{\textrm{IM}}$ denote the modulation amplitude of the real and the imaginary part of relative permittivity, respectively.}
	\label{fig:geom2}
\end{figure}

In the main text of the manuscript, we have studied the nonlinear $\mathcal{PT}$-symmetric coupler built of three dimers. Here, we compare its dispersion relations and the nonlinear dynamics of light propagation with simpler structures built of one or two dimers described by \cref{eqn:cos_ind} in the main text. The geometries of the structures described in these supplementary materials are presented in \cref{fig:geom2}.
 
\subsubsection{Linear dispersion diagrams for a single multimode dimer}


\begin{figure}[!b]
	\includegraphics[width = 0.49\columnwidth, clip=true, trim = {0 0 10 0}]{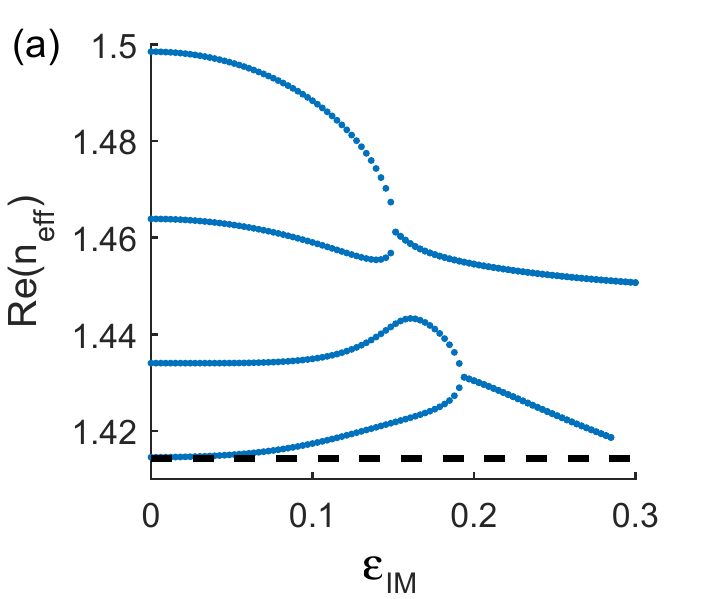}
	\includegraphics[width = 0.49\columnwidth, clip=true, trim = {0 0 10 0}]{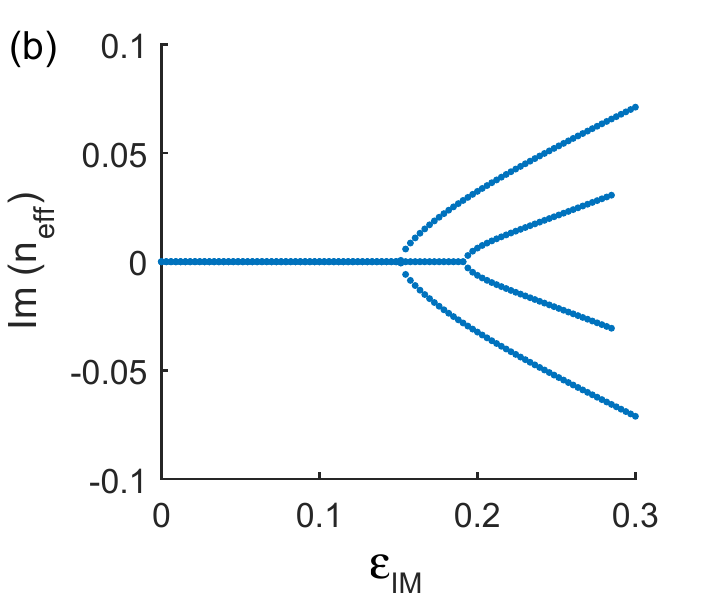}
	\caption{Linear dispersion relations for a single cosine-like dimer described by \cref{eqn:cos_ind} and shown in \cref{fig:geom2}(a). The (a) real part and (b) imaginary part of the effective index are shown as a function of the imaginary part of the permittivity modulation depth $\epsilon_{\textrm{IM}}$. Dimer parameters (given in the text) are chosen in such a way that the dimer supports multiple modes. The black dashed line denotes the refractive index of a background medium $\sqrt{\epsilon_B}$.}
	\label{fig:single_per_disp_high}
\end{figure}

In the main text, we have reported a new type of dispersion diagrams for the multimode couplers built of more than one dimer. Here, we compare these new dispersion diagrams [shown in~\cref{fig:disp_multi3}(a),~(b) in the main text] with the dispersion relations of a multimode waveguide composed of a single $\mathcal{PT}$-symmetric dimer. \Cref{fig:single_per_disp_high} presents the dispersion relations $n_{\textrm{eff}}(\epsilon_{\textrm{IM}})$ for a single dimer shown in \cref{fig:geom2}(a). The parameters of the dimer are $\Gamma = 3$~$\mu$m, $\epsilon_B = 2$, and $\epsilon_{\textrm{RE}} = 0.3$. The modulation depth of the real part of permittivity is chosen in such a way that the dimer supports more than one pair of modes. In this case, the dispersion curves of higher-order modes appear above the dispersion curves of the lowest-order modes, as it was shown in Ref.~\cite{Huang14}. In contrast to the dispersion relation of a single multimode dimer presented in~\cref{fig:single_per_disp_high}, the dispersion relations of a multimode coupler that is built of more than one dimer, as shown in~\cref{fig:disp_multi3} in the main text and in~\cref{fig:two_per_disp}, present a different type of behavior. In this system, the dispersion curves of the higher-order modes appear inside the curves corresponding to the lower-order modes.


\subsubsection{Coupler built of two dimers}

\begin{figure}[!b]
	\includegraphics[width = 0.49\columnwidth, clip=true, trim = {0 0 10 0}]{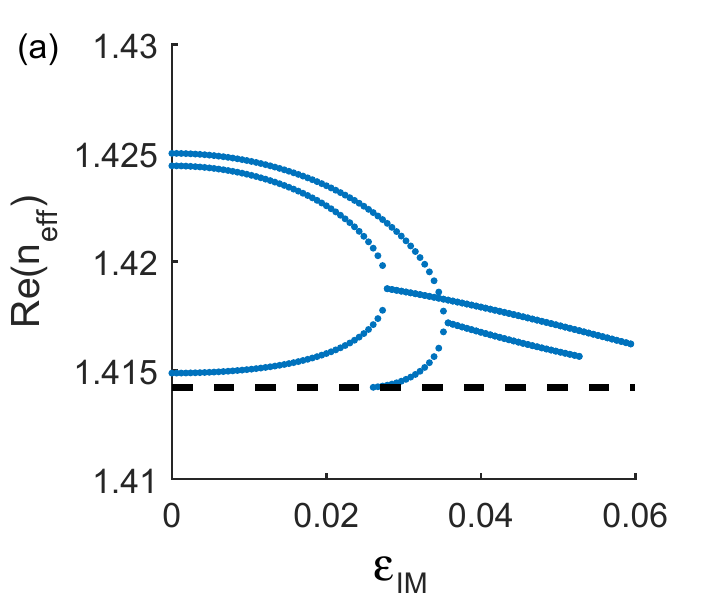}
	\includegraphics[width = 0.49\columnwidth, clip=true, trim = {0 0 10 0}]{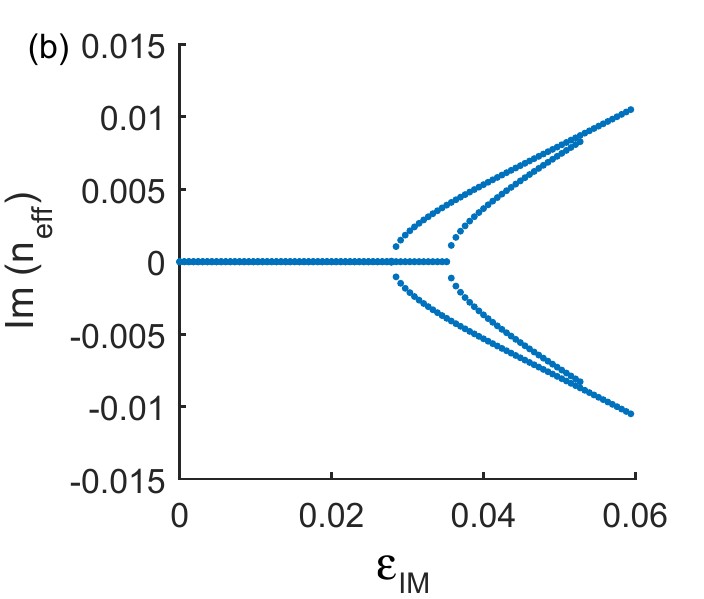}
	\includegraphics[width = 0.49\columnwidth, clip=true, trim = {205 310 215 315}]{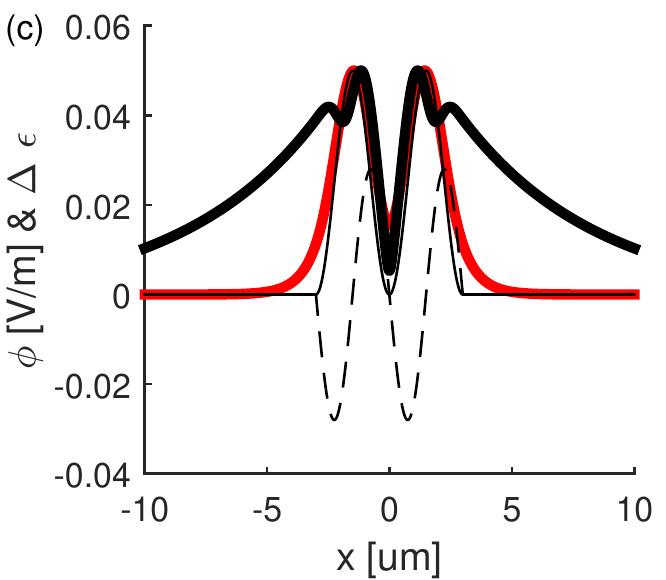}
	\includegraphics[width = 0.49\columnwidth, clip=true, trim = {205 310 215 315}]{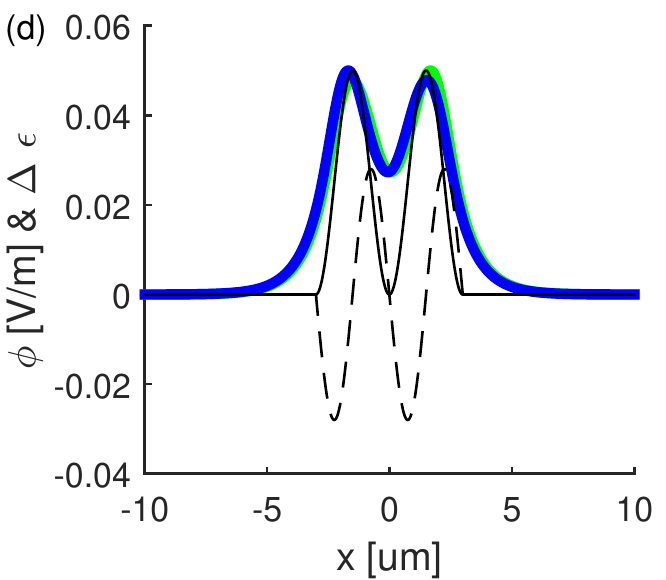}
	\caption{(a),~(b) Dispersion relations for a $\mathcal{PT}$-symmetric coupler composed of two dimers; (a) the real and (b) the imaginary part of the effective index of the modes as a function of the imaginary part of permittivity $\epsilon_{\textrm{IM}}$ in the linear case. The black dashed line denotes the refractive index of a background medium $\sqrt{\epsilon_B}$. (c),~(d) Absolute value of the field distributions $|\phi(x)|$ of (c) the lossless modes [mode~1~(fundamental)---red, mode 4 (lowest effective index)---black], (d) the coupled modes with loss (blue) and gain (green) at $\epsilon_{\textrm{IM}} = 0.028$. Black curves indicate $\Re e\{\Delta \epsilon\}$ (solid) and $\Im m\{\Delta \epsilon\}$ (dashed).}
	\label{fig:two_per_disp}
\end{figure}

In the main text, we have presented the description of a nonlinear coupler described by a cosine-like permittivity profile [\cref{eqn:cos_ind} in the main text] built of three $\mathcal{PT}$-symmetric dimers. Here, we complete the description by studying an array composed of two dimers, whose geometry is presented in~\cref{fig:geom2}(b). The linear dispersion relations of the system of two dimers, presented in~\cref{fig:two_per_disp}(a),~(b), show qualitatively similar behavior to the coupler built of three dimers. Here, the novel type of dispersion diagrams is also observed, where the dispersion curves of the higher-order mode pairs enclose the curves corresponding to lower-order mode pairs. This confirms our conclusions that the dispersion curves of a multimode $\mathcal{PT}$-symmetric coupler built of a single dimer are qualitatively different from the curves of couplers built of more than one dimer.

\begin{figure}[!t]
	\includegraphics[width = 0.49\columnwidth, clip=true, trim = {0 0 20 0}]{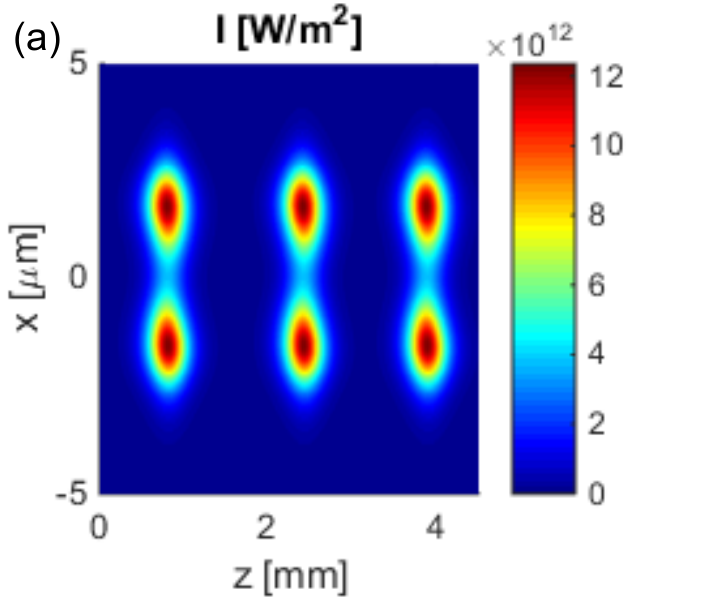}
	\includegraphics[width = 0.49\columnwidth, clip=true, trim = {0 0 20 0}]{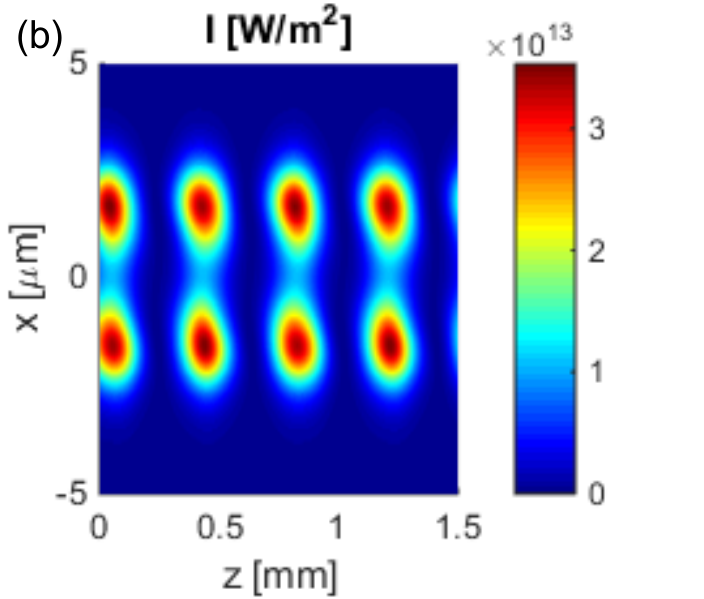}
	\includegraphics[width = 0.49\columnwidth, clip=true, trim = {0 0 20 0}]{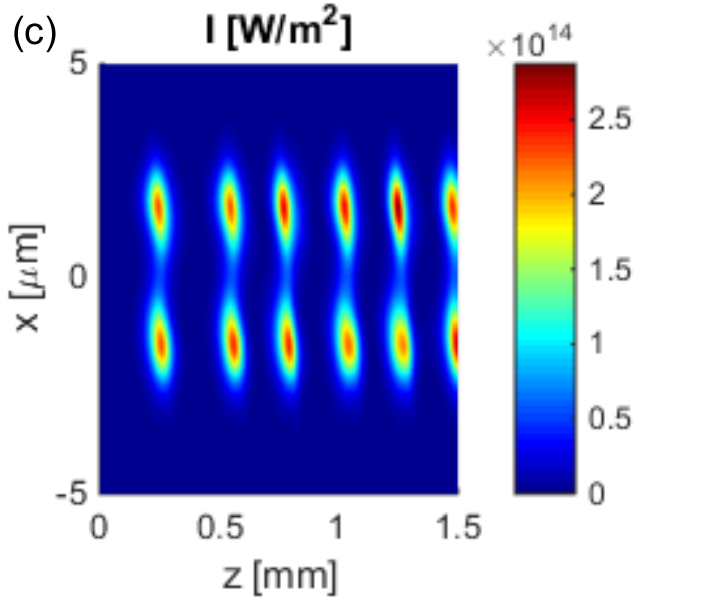}
	\includegraphics[width = 0.49\columnwidth, clip=true, trim = {0 0 20 0}]{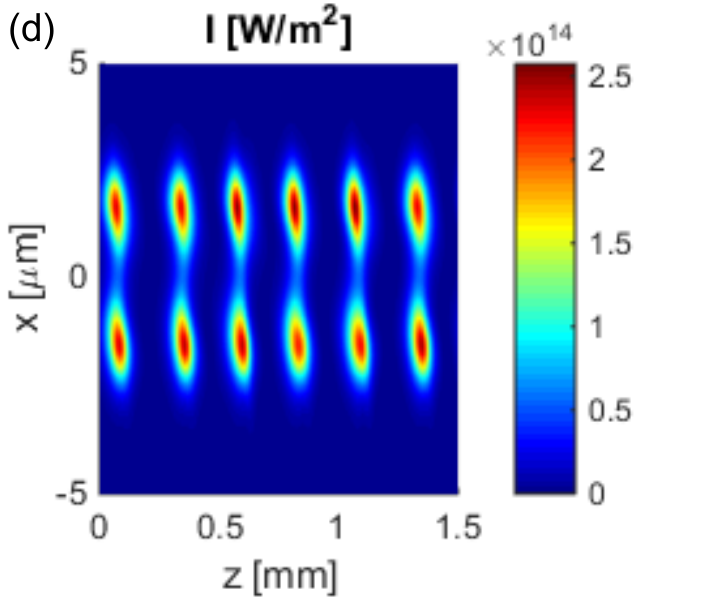}
	\caption{The intensity distributions $I(x,z)$ showing the nonlinear dynamics of light propagating in an array built of two dimers described by \cref{eqn:cos_ind}. The geometry of the structure is shown in \cref{fig:geom2} and in the inset of supblot (a). The parameters are: $\Gamma = 3$~$\mu$m, $\epsilon_{\textrm{RE}} = 0.05$, $\alpha = 10^{-19}$~m$^2$/V$^2$ and (a),~(b) $\epsilon_{\textrm{IM}} = 0.028$, (c),~(d) $\epsilon_{\textrm{IM}} = 0.030$. Excitation power density $P_0$  is equal to: (a),~(c) $10^5$~W/m and (b),~(d) $10^8$~W/m. The propagation distance shown in (a) is three times longer than in (b)--(d).}
	\label{fig:two_per_evo}
\end{figure}

Similar to the case of a coupler built of three dimers, here we also observe the nonlinearly-triggered-$\mathcal{PT}$~transition from the full to the broken $\mathcal{PT}$-symmetric regime. Linear modal studies of this system  show that this transition occurs at $\epsilon_{\textrm{IM}} = 0.0278$. The nonlinear propagation of light in the system just above the $\mathcal{PT}$~transition at low and high input power levels is shown in Figs.~\ref{fig:two_per_evo}(a) and (b), respectively. At the low power level, the oscillation period is approximately equal to $1.4$~mm, and the peak power corresponds to the nonlinear permittivity modulation depth of the order of $6\cdot10^{-4}$. For high power, the period decreases to $0.4$~mm, and the maximum nonlinear permittivity modulation depth increases to $1.8\cdot10^{-3}$. In the case of two dimers, the effect of the increase of the amplitude of the imaginary part of permittivity is also studied. Figures~\ref{fig:two_per_evo}(c),~(d) present the light propagation for the same two input powers as in Figs.~\ref{fig:two_per_evo}(a),~(b) but in the system located higher above the $\mathcal{PT}$~transition. In this case, the oscillation period decreased to $0.25$~mm.
Contrary to the case of lower amplitude of the imaginary part, the increase of the initial power for the case presented in Figs.~\ref{fig:two_per_evo}(c) and (d) does not modify the oscillation period, and it is the same for both power levels. Only the initial propagation distance required to initiate the oscillations (from the input plane $z=0$ to the first intensity maximum) is reduced with the increase of the excitation power because the same intensity level is reached earlier by the system, where more power was injected.

\begin{figure}[!h]
	\includegraphics[width = \columnwidth, clip=true, trim = {120 310 145 310}]{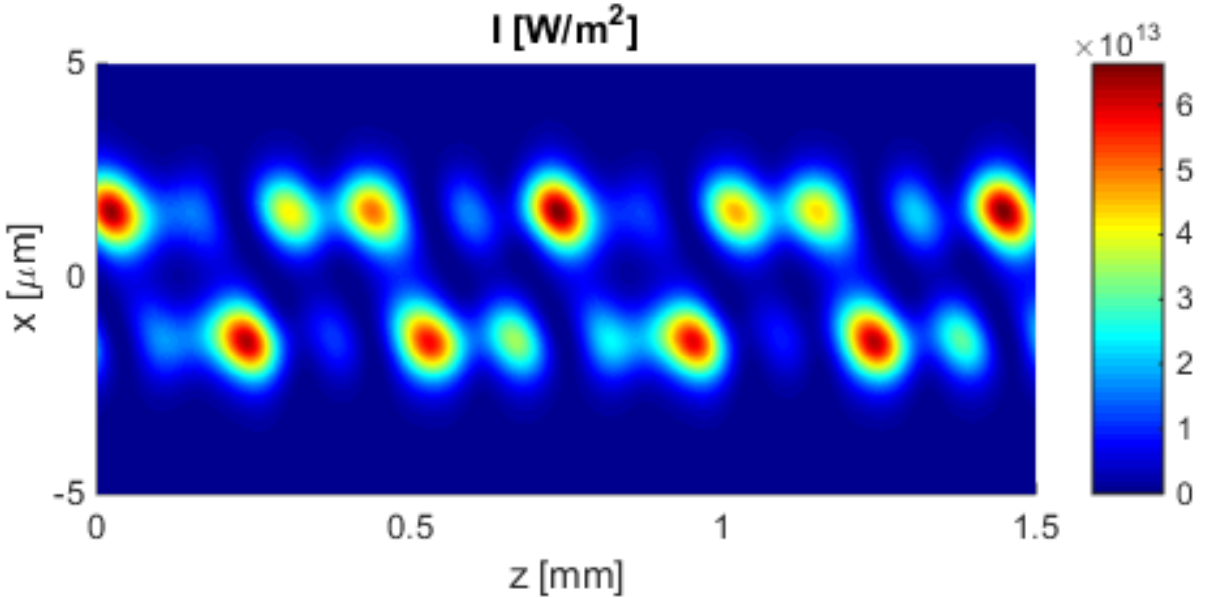}
	\caption{The intensity distributions $I(x,z)$ showing the nonlinear dynamics of light propagating in an array built of two dimers with the same parameters as these presented in \cref{fig:two_per_evo}(b). The input field is in the form of the sum of mode~1 and gain mode. The excitation power density is $P_0 = 10^8$~W/m [the same as in~\cref{fig:two_per_evo}(b)].}
	\label{fig:two_per_evo_mode12}
\end{figure}

Finally, we have studied the behavior of the coupler built of two $\mathcal{PT}$-symmetric dimers when more than one mode of the system is excited. The field profiles of the linear modes supported by the system are presented in~\cref{fig:two_per_disp}(c),~(d). \Cref{fig:two_per_evo_mode12} presents the evolution of the field in the case when the gain mode and the fundamental mode are excited simultaneously. The interference pattern obtained here resembles these for the coupler built of three waveguides presented in~\cref{fig:three_per_evo_mode12} in the main text. However, for the simpler system composed of two waveguides, the interference pattern is less complex than for the coupler built of three $\mathcal{PT}$-symmetric waveguides.

\end{document}